\documentclass[useAMS,usenatbib]{mn2e}
\usepackage{graphicx}
\usepackage{multirow}
\usepackage[centertags]{amsmath}
\usepackage{color}
\usepackage{amsfonts}
\usepackage{amssymb}

\title[Seismic modelling of the SPB star HD\,21071]{Seismic modelling of the rotating, slowly pulsating B-type star HD\,21071}

\author[W. Szewczuk and J. Daszy\'nska-Daszkiewicz]{W. Szewczuk$^{1}$\thanks{E-mail:
szewczuk@astro.uni.wroc.pl (WS);\newline daszynska@astro.uni.wroc.pl (JDD)} and
J. Daszy\'nska-Daszkiewicz$^{1}$\footnotemark[1]{}\\
$^{1}$Instytut Astronomiczny, Uniwersytet Wroc{\l}awski,
    Kopernika 11, 51-622 Wroc{\l}aw, Poland}

\begin{document}

\date{Accepted 1988 December 15. Received 1988 December 14; in original form 1988 October 11}

\pagerange{\pageref{firstpage}--\pageref{lastpage}} \pubyear{2002}

\maketitle

\label{firstpage}

\begin{abstract}
Interpretation of the oscillation spectrum of the slowly pulsating B-type star HD21071 is presented.
We show that non-rotating models cannot account for the two highest amplitude frequencies
and taking into account the effects of rotation is necessary.
Rotating seismic models are constructed using various chemical compositions, opacity data, core overshooting parameters and rotational velocities.
There are prospects for seismic modelling of SPB stars, even if no asymptotic pattern is observed in their oscillation spectra,
provided an unambiguous mode identification is doable and the effects of rotation are properly included.
\end{abstract}


\section{Introduction}

Main sequence stars of mid to late B spectral types pulsating with periods
of the order of days have been dubbed slowly pulsating B-type stars (SPB stars) by
\citet{Waelkens1991}. Although multiperiodicity and long periods
undoubtedly pointed to pulsations in high order gravity modes,
the excitation mechanism remained unknown at that time.

Progress has been made possible soon after publication of the revised opacity data
OPAL \citep*{Iglesias1992} with a new maximum around $\log T\approx5.3$
(the so-called metal or $Z$ opacity bump) caused by numerous absorption lines of iron-group elements.
Using the new opacity data, \citet*{DMP1993} and \citet*{Gautschy1993} showed that high radial order gravity modes
in SPB stars are driven by the classical $\kappa$-mechanism operating in the metal opacity bump.

However, studying SPB stars is not an easy task from both observational and theoretical points of view.
Observations of SPB stars are challenging because their low frequencies
demand long runs of observations. Furthermore, extracting periods of the order of a day
is often complicated because of an aliasing effect, especially from ground-based observations.
Therefore until the Hipparcos mission only a few SPB stars were known.
Thanks to the Hipparcos photometry, \citet{Waelkens1998} increased the number of SPB stars from 11 to about 100.
Since the pulsation periods are often of the same order as the rotation periods, the next obstacle
is to distinguish them from each other. In many cases this task is impossible without detailed spectroscopic analysis.

Theoretical interpretation of the SPB pulsations is complicated because of three reasons:
mode identification, very dense theoretical oscillation spectra and rotation.
Firstly, because of the lack of clear structures in the observed oscillation spectra,
the only way to determine the spherical harmonic degree, $\ell$,
and the azimuthal order, $m$, is the use of the information
contained in the light and line profile variations
\citep*[e.\,g.][]{Balona1979, Watson1988,Cugier1994,Campos1980a,Campos1980b,
Balona1986a,Balona1986b,CJDD2001,Gies1988,Kennelly1996,Telting1997}.
An exception are the most recent observational results from the CoRoT
and Kepler space missions \citep[e.g.,][]{Degroote2010,Papics2012,Papics2014,Papics2015}.
However, we need to emphasize that in very dense oscillation spectra such structures can be accidental.
An example is the SPB star HD\,50230. \citet{Degroote2010} claim that in the oscillation
spectrum of the star there is present a sequence equally-spaced in period associated with
asymptotic properties while our studies \citep*{WSJDDWD2014} contradict this.
Secondly, because of very dense theoretical oscillation spectra of SPB stellar models,
assignment of the radial order, $n$, to the individual observed peaks is usually very ambiguous.
The last problem concerns the effects of rotation, which influence, both, the equilibrium model and pulsational properties.
In the case of SPB stars, even at slow rotation, pulsational frequencies can be of the order
of the rotational frequency and the effects of the Coriolis force cannot be neglected.
On the other hand, the advantage is that the influence of centrifugal deformation on low
frequency gravity modes is small \citep{Ballot2012}.
Effects  of rotation on low frequency  g modes was studied in the framework
of the so-called traditional approximation
\citep*[e.\,g.,][]{Lee1997, Townsend2003a, Townsend2003b, Townsend2005, JDD2007}
or using the truncated expansion for the eigenfunctions \citep[e.\,g.,][]{Lee1989, Lee2001}.

Attempts to match the observed frequencies to the theoretical ones have been undertaken
by \citet*{Walczak2012, Walczak2013}, who found seismic models which fit two observed frequencies
with well identified degrees, $\ell$, in the two SPB stars HD\,74560 and HD\,182255.
However, they assumed that the observed frequencies are the axisymmetric modes, i.\,e., $m=0$,
and applied the zero-rotation approximation in pulsational calculations.

Without a doubt, the best example of seismic modelling of the SPB stars so far is that of KIC\,10526294.
Using the Kepler time-series photometry, \citet{Papics2014} found 19 frequency peaks quasi-equally spaced in period,
most of which were split into triplets. These authors interpreted the 19 central peaks
according to the asymptotic theory as dipole g modes with consecutive high radial orders, $n$.
However, KIC\,10526294 is unique among the SPB stars because of its extremely slow rotation.
The rotational period, as deduced from the rotationally split modes, is equal to 188 days,
which justified neglecting the effects of rotation in seismic modelling.

Furthermore, \citet{Papics2015} found 36 frequency peaks quasi-equally spaced in period
being probably a manifestation of the asymptotic properties in the SPB star KIC\,776080. The star
seems to be even more attractive than KIC\,10526294 not only because of the longer equally spaced modes series
but also because of the higher rotation velocity of at least 62 km s$^{-1}$.
Detailed seismic modelling is still to be done.

The goal of this paper is to present results of seismic modelling of the SPB star HD\,21071.
The star has a few frequency peaks determined from ground-based photometry with a well identified angular numbers ($\ell, m$)
\citep{WSJDD2015MNRAS}.
The effects of rotation are included via the traditional approximation.
In the next section, we present the star and results of earlier studies.
In Section 3, the  oscillation spectrum of the star is interpreted.
Section 4 is devoted to seismic modelling.
In the last section we summarize the results and discuss the prospect for future studies.

\section{The SPB star HD\,21071}
HD\,21071 (HR\,1029, HIP\,15988, V576\,Per) is a star of brightness
$V=6.1$ mag \citep{Reed2005} and spectral type B7V
\citep*{1971AJ.....76..242M,2013ApJ...771..110M}.
Slightly different spectral classifications can be also found in the literature,
e.\,g., B6V \citep*{1962ApJ...136..381A}, B5IV \citep*{1968ApJS...17..371L} and B4/B3  \citep*{2012AJ....143...28F}.

\begin{figure}
\centering
\includegraphics[angle=-90, width=\columnwidth]{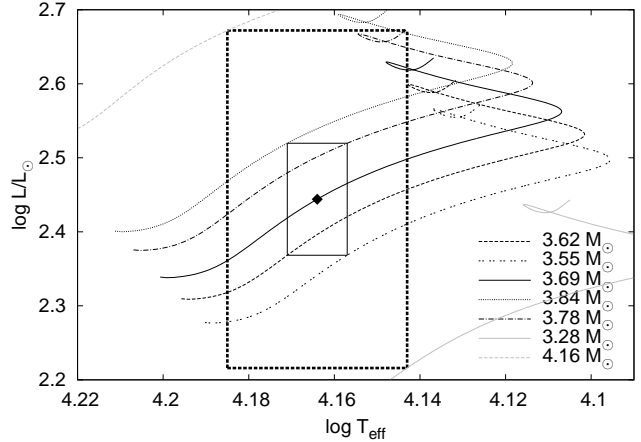}
\caption{The position of HD\,21071 (black diamond) in the HR diagram with $1\sigma$ error box (solid line rectangle)
            and $3\sigma$ error box (dash-dotted line rectangle).
           Also shown are evolutionary tracks passing through the centre and edges
           of the error boxes. Evolutionary  calculations were performed using
           the Warsaw-New Jersey evolutionary code, OP opacities,
           AGSS09 chemical mixture, the initial hydrogen abundance  $X_0=0.7$,
           metallicity $Z=0.0082$ and without overshooting from the convective core.
           The rotational velocity at the ZAMS was $\mathrm V_\mathrm{rot}=50$ km s$^{-1}$.
            \label{HR}}
\end{figure}

There are many determinations of the effective temperature
from photometric calibrations as well as spectral fitting, e.\,g.,
$\log T_\mathrm{eff}=4.169$ \citep*{2014A&A...562A.128S},
$\log T_\mathrm{eff}=4.149$ \citep*{2014A&A...566A.132S},
$\log T_\mathrm{eff}=4.212$ \citep*{2012AJ....143...28F},
$\log T_\mathrm{eff}=4.157$ \citep*{2012A&A...537A.120Z},
$\log T_\mathrm{eff}=4.130$ \citep*{2010A&A...515A..74L}.
In this paper we adopted $\log T_\mathrm{eff}=4.164\pm0.007$ obtained
by \citet{Niemczura2003} from the IUE ultraviolet spectra.
This value is approximately equal to the mean of the values obtained
by the above mentioned authors. The metallicity, as determined from the IAU spectra, is $0.0082^{+0.0053}_{-0.0032}$
\citep{Niemczura2003}.
To put the star on the HR diagram presented
in Fig.\,\ref{HR}, we used $\log  L/\mathrm L_{\sun}=2.444\pm0.076$ derived by \citet{WSJDD2015MNRAS}.

In Fig.\,\ref{HR}, there are also shown the evolutionary tracks calculated
with the OP opacity tables \citep{Seaton2005}
and the latest heavy element mixture of \citet{Asplund2009} (hereafter AGSS09).
We assumed the initial hydrogen abundance $X_0=0.7$, the metallicity $Z=0.0082$
and the equatorial rotational velocities on the ZAMS equal to the projected value,
$\mathrm{V_{rot}} \sin i=50$ km s$^{-1}$, obtained by \citet*{Abt2002}.
Overshooting from convective core was not taken into account.
The Warsaw-New Jersey code \citep[e.\,g.][]{Pamyatnykh1998}  is used in the evolutionary calculations throughout the paper.
The estimated evolutionary mass for HD\,21071 is $M=3.69$ M$_{\sun}$
and the ranges of masses for the $1\sigma$ and $3\sigma$ error box
are $\left<3.55,~3.84\right>$ and $\left<3.28,~4.16\right>$, respectively.

\section{Oscillation frequencies of HD\,21071}

HD\,21071 was classified as an SPB star by \citet{Waelkens1998} who found
variability with a period of $P=0.84$ d in the Hipparcos space photometric data.
In the seven band of the Geneva photometric system, \citet{DeCat2007} detected four frequencies:
$\nu_1=1.18843\,\mathrm d^{-1}$, which coincidences with this given by \citet{Waelkens1998},
$\nu_2=1.14934\,\mathrm d^{-1}$ (also present in the Hipparcos data),
$\nu_3=1.41968\,\mathrm d^{-1}$ and $\nu_4=0.95706\,\mathrm d^{-1}$.
In Table\,\ref{freq_ident}, we give the values of these frequencies and the corresponding amplitudes
in the $UBV$ Geneva filters (columns from 2 to 5).
It should be mentioned that due to a strong aliasing there is a risk that
$\nu_3$ and $\nu_4$ can be mistaken with their aliases.
Two dominant frequencies, $\nu_1$ and $\nu_2$, were also detected
by \citet{Andrews2003} in the APT data.
Moreover $\nu_1$ was detected in spectroscopic data \citep{DeCat2002ASPC_259_A}.

\begin{table*}
 \centering
 \begin{minipage}{\textwidth}
  \caption{Frequencies detected in HD\,21071 with the corresponding amplitudes of the light (in the Geneva $UBV$ filters) and radial velocity
               variations. The last three columns contain the results of the mode identifications.}
\label{freq_ident}
\centering
  \begin{tabular}{lcccccccc}
  \hline
ID     &   $\nu^a$          & A$_U^a$    & A$_B^a$    & A$_V^a$    & A$_{V_\mathrm{rad}}^b$    & $\ell^a$  & $\ell^c$     &   $\left( \ell,\,m\right)^c$\\
       &$(\mathrm d^{-1})$& (mmag) & (mmag) & (mmag) & (km s$^{-1}$) & \multicolumn{2}{|c|}{$\mathrm{V_{rot}}=0$ (km s$^{-1}$)}  & $\mathrm{V_{rot}}\in \left<150,\,250\right>$ (km s$^{-1}$)\\
\hline
\hline
$\nu_1$&  1.18843       & 34.3     & 21.3     & 18.5     & 3.27          & 1       & 1     &  $(1,\,0)$\\
       & $\pm$0.00001   & $\pm$0.8 & $\pm$0.6 & $\pm$0.6 & $\pm$0.85     &         &       &  \\
\hline
$\nu_2$&   1.14934      & 15.4     & 8.8      & 7.7      &               & \,1; 2; 4\, & \,\,\,\,1; 2\,\,\,\,  &  $(1,\,0)$\\
       & $\pm$0.00003   & $\pm$0.8 & $\pm$0.7 & $\pm$0.6 &               &         &       &  \\
\hline
$\nu_3$& 1.41968        & 6.7      & 4.1      & 3.8      &               & 1; 2    & 1; 2  &  $(1,\,0)$\\
       & $\pm$0.00007   & $\pm$0.8 & $\pm$0.6 & $\pm$0.6 &               &         &       &  \\
\hline
$\nu_4$&   0.95706      & 5.8      & 3.3      & 3.0      &               & 1; 2; 4 & 1; 2  &  $(1,\,0)$; $(1,\,-1)$;\\
       & $\pm$0.00009   & $\pm$0.8 & $\pm$0.6 & $\pm$0.6 &               &         &       &  $(2,\,-1)$; $(2,\,-2)$ \\
\hline
\footnotetext{
$^a$\citet{DeCat2007};
$^b$\citet{DeCat2002ASPC_259_A};
$^c$\citet{WSJDD2015MNRAS}
}
\end{tabular}
\end{minipage}
\end{table*}

\begin{figure}
\centering
\includegraphics[angle=-90, width=\columnwidth]{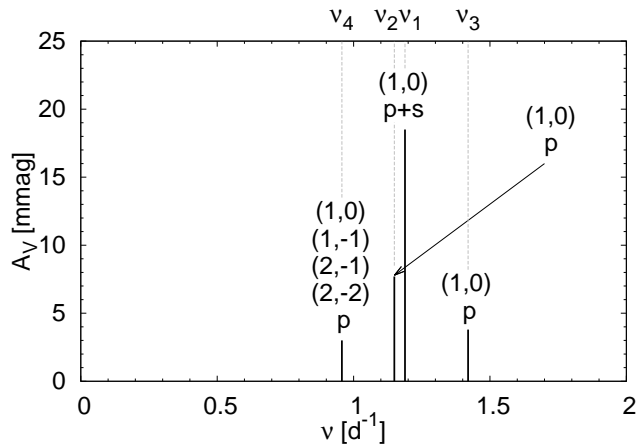}
\caption{Oscillation spectrum of HD\,21071. In parentheses are given $\ell$ and $m$,
         while 'p' and 's' indicate the frequency observed in photometry and spectroscopy, respectively.
            \label{osc_spec}}
\end{figure}

Identification of the degree, $\ell$, for the frequencies of HD\,21071
was performed by \citet{DeCat2007} and the results are given in the 7th column of Table\,\ref{freq_ident}.
Recently, \citet{Szewczuk2015,WSJDD2015MNRAS} included the effects of the Coriolis force
in the mode identification. They determined both angular numbers $(\ell, m)$
as well as constrained the rotational velocity ($\mathrm{V_{rot}}\in \left<150,\,250\right>$ [km s$^{-1}$]).
The values of $(\ell, m)$ are given in the last column of Table\,\ref{freq_ident}.
For completeness, in the penultimate column we added identification of $\ell$
obtained by \citet{Szewczuk2015,WSJDD2015MNRAS} with the zero-rotation approximation.

As one can see from Fig.\,\ref{osc_spec}, three frequencies HD\,21071, i.\,e. $\nu_1$, $\nu_2$ and $\nu_3$,
are dipole axisymmetric modes. The frequency $\nu_4$ can be a dipole axisymmetric or a retrograde mode,
or a quadrupole retrograde mode.
One can also see that $\nu_4$, $\nu_1$ and $\nu_3$ are equally spaced
in frequency. If we are not in the asymptotic
regime of high order acoustic modes, which is obviously true in our case,
a rotational origin of triplets is a most probable explanation for such type of structures.
If we are dealing with a rotationally
split triplet, the central peak, $\nu_1$, should be an axisymmetric mode, the left side
peak, $\nu_4$, should be a retrograde mode and the right side peak, $\nu_3$, a prograde mode.
However, this is in contradiction with the identification of $\nu_3$ as an axisymmetric mode.

Another evidence suggesting the non-rotational origin of the triplet is its perfect symmetry.
The difference between the central and left side peaks,
$\nu_1-\nu_4=0.23137\pm0,00009$,  is equal within the errors to the difference
between the right side and the central peaks, $\nu_3-\nu_1=0.23125\pm0,00007$.
Linear pulsational theory predicts an asymmetry and the
left side peak should be closer to the central one than the right side peak
(see also Appendix A for more details, only  in the electronic edition of the journal).

Because the study of higher order effects of mode coupling is beyond the scope of this paper,
we assumed that the triplet in the oscillation spectrum of HD\,21071 is accidental.

\section{Seismic modelling}

\subsection{Zero rotation-approximation}

\begin{figure}
\centering
\includegraphics[angle=-90, width=\columnwidth]{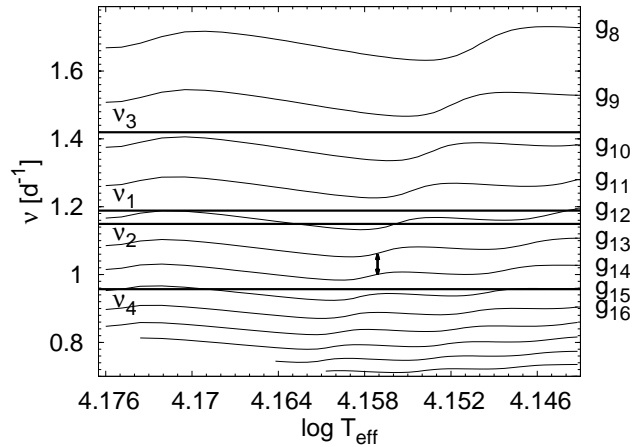}
\caption{Theoretical frequencies of dipole modes as a function of the effective temperature.
            The frequencies were calculated in the zero-rotation approximation for the model with the parameters
            $M=3.52 \mathrm  M_{\sun}$, $\log T_\mathrm{eff}=4.1571$, $\log L/\mathrm L_{\sun}=2.277$
            and $Z=0.010$. The smallest separation between consecutive radial orders
            with frequencies above 1 d$^{-1}$ is indicated by the two headed arrow.
            The observed frequencies are marked by thick horizontal lines.
            \label{zero_rot}}

\end{figure}

The two highest amplitude frequencies, $\nu_1$ and $\nu_2$, are separated from each other
by only 0.03909 d$^{-1}$. Since from the mode identification we know they are dipole axisymmetric
modes, such small separation can suggest that they are consecutive radial orders.
To check this hypothesis, we constructed a preliminary grid of pulsational models in
the framework of the zero-rotation approximation.
We calculated models lying within the $3\sigma$ error box (see Fig.\,\ref{HR})
for three values of metallicities, $Z=0.010,\, 0.015,\, 0.020$, the initial hydrogen
abundances, $X_0=0.7$, OP opacity tables and AGSS09 chemical mixture with a step
in mass of 0.005 $\mathrm M_{\sun}$. No overshooting from convective core has been taken into account.
It turned out that a minimal separation between consecutive radial orders of dipole modes
in the frequency range appropriate for $\nu_1$ and $\nu_2$ (i.e. above 1 d$^{-1}$)
is $min\left\{\nu(\mathrm g_n)-\nu(\mathrm g_{n+1})\right\}=0.0618$ d$^{-1}$.
This minimum theoretical separation is almost two times larger than the observed difference
between $\nu_1$ and $\nu_2$ and was achieved between modes $\mathrm g_{13}$
and $\mathrm g_{14}$ in the model with $M=3.52 ~\mathrm M_{\sun}$, $Z=0.010$, $\log T_\mathrm{eff}=4.1571$
and $\log L/\mathrm L_{\sun}=2.277$.
In Fig.\,\ref{zero_rot}, there is shown the evolution of the frequencies of dipole modes
from ZAMS to the rightmost edge of the $3\sigma$ error box. As one would expect,
$min\left\{\nu(\mathrm g_n)-\nu(\mathrm g_{n+1})\right\}$ (marked by the two headed arrow) coincides with
the avoided crossing phenomenon.

As a consequence, to properly interpret the frequencies $\nu_1$ and $\nu_2$,
one has to include effects of rotation. In the rotating star the eigenvalue
$\ell\left(\ell+1\right)$ is replaced by $\lambda$
which for axisymmetric modes is increasing function of $\mathrm{V_{rot}}$.
Increasing $\lambda$ imitates the effect of the higher degree, $\ell$,
and, as a consequence, the oscillation spectrum is more dense for a given frequency range.

\subsection{Including rotation}
In this section, we construct seismic models fitting the two ($\nu_1$, $\nu_2$)
and three ($\nu_1$, $\nu_2$, $\nu_3$) observed frequencies of HD\,21071.
The effects of rotation were included via the traditional approximation in
which the effects of the Coriolis force are taken into account
whereas centrifugal distortion is neglected. Rigid rotation is assumed.

The method of mode identification developed by \citet{JDD2015}
allows, besides determination of $\ell$ and $m$, to constrain the rotational velocity.
\citet{WSJDD2015MNRAS} showed that the rotation rate of HD\,21071 is in the range $\mathrm{V_{rot}}\in\left<150,\,250\right>$ km s$^{-1}$.
We confine our searching of seismic models to this range of the rotational velocity.
Due to a high computational cost, it was done in the two steps.

Firstly, we constructed a preliminary grid of models for $X_0=0.7$,
with the OP opacity data, AGSS09 chemical mixture and without
overshooting from the convective core ($\alpha_{\rm ov}=0.0$).
Models were calculated inside the $3\sigma$ error box (see Fig.\,\ref{HR})
for different masses, metallicities and rotation velocities with a step
$\Delta {M}=0.01~\mathrm M_{\sun}$, $\Delta Z=0.00025$ and $\Delta \mathrm{V_{rot}}=10$ km s$^{-1}$.
Then, to find the approximate parameters of models which fit the two observed frequencies,
the theoretical frequencies were interpolated by means of multi-dimensional linear interpolation.

In the second step, around the approximate model parameters for which
theoretical frequencies fit the observed ones, we computed a denser
grid of models with steps $\Delta {M}=0.002~ \mathrm M_{\sun}$ and $\Delta Z=0.00005$.
A step in effective temperature in both grids was dynamically determined by the evolutionary code
and equal approximately to $\Delta T_\mathrm{eff} \simeq 0.0005$.

We will call the grid of models with the input ($X_0=0.7$, OP, AGSS09, $\alpha_{\rm ov}=0.0$) as G1.

\begin{table*}
\centering
\begin{minipage}{\textwidth}

\caption{{The radial orders, $n$, of dipole modes which fit the three frequencies, $\nu_1$, $\nu_2$ and $\nu_3$, of HD\,21071
  in the models G1, G2, G3 and G4. In square brackets are stable modes. The first column contains the value of the rotational velocity.}}
\label{radial_orders_v1v2v3}
\centering
  \begin{tabular}{ccccc}
    \hline
     &  {\bf G1} & {\bf G2} & {\bf G3} &{\bf G4} \\
  \hline
$\mathrm{V_\mathrm{rot}}$                  & OP               &             OP    & OP                & OPAL  \\
$\left( \mathrm{km\,s}^{-1}\right)$ & $X_0=0.70$         & $X_0=0.75$          & $X_0=0.70$          & $X_0=0.70$  \\
                                    & $\alpha_\mathrm{ov}=0.0$& $\alpha_\mathrm{ov}=0.0$ & $\alpha_\mathrm{ov}=0.2$ & $\alpha_\mathrm{ov}=0.0$ \\

\hline\hline

150 & $[g_{16},\,g_{17},\,g_{12}]$  &$[g_{16},\,g_{17},\,g_{12}]$& $g_{14},\,g_{15},\,g_{10}$  & $[g_{16},\,g_{17},\,g_{12}]$   \\
    & $g_{14},\,g_{15},\,g_{10}$    &                            &                             & $[g_{15},\,g_{16},\,g_{11}]$   \\
    &                               &                            &                             & $g_{14},\,g_{15},\,g_{10}$   \\
\hline
160 & $g_{13},\,g_{14},\,g_{9}$     & $g_{13},\,g_{14},\,g_{9}$  & $g_{14},\,g_{15},\,g_{10}$  & -                            \\
    &                               &                            & $g_{13},\,g_{14},\,g_{9}$   &                              \\
\hline
170 & $g_{14},\,g_{15},\,g_{10}$    & $g_{14},\,g_{15},\,g_{10}$ &$[g_{14},\,g_{15},\,g_{10}]$ & -                            \\
\hline
180 & $g_{14},\,g_{15},\,g_{10}$    &$[g_{15},\,g_{16},\,g_{11}]$&   -                         & $[g_{29},\,g_{31},\,g_{20}]$   \\
    &                               & $g_{14},\,g_{15},\,g_{10}$ &                             & $g_{14},\,g_{15},\,g_{10}$   \\
\hline
190 &       -                       &     -                      &   -                         & $[g_{14},\,g_{15},\,g_{10}]$   \\
\hline
200 &       -                       &     -                      &   -                         & -                            \\
\hline
210 & $[g_{29},\,g_{31},\,g_{20}]$  &     -                      &   -                         & -                            \\
\hline
220 & $[g_{29},\,g_{31},\,g_{20}]$  &$[g_{29},\,g_{31},\,g_{20}]$&   -                         & -                            \\
\hline
230 &       -                       &     -                      &   -                         & $[g_{29},\,g_{31},\,g_{20}]$   \\
\hline
240 &       -                       &     -                      &   -                         & -                            \\
\hline
250 &       -                       &     -                      &   -                         & $[g_{28},\,g_{30},\,g_{19}]$   \\
\hline
\end{tabular}
\end{minipage}
\end{table*}

\begin{figure*}
\centering
\includegraphics[angle=-90, width=2\columnwidth]{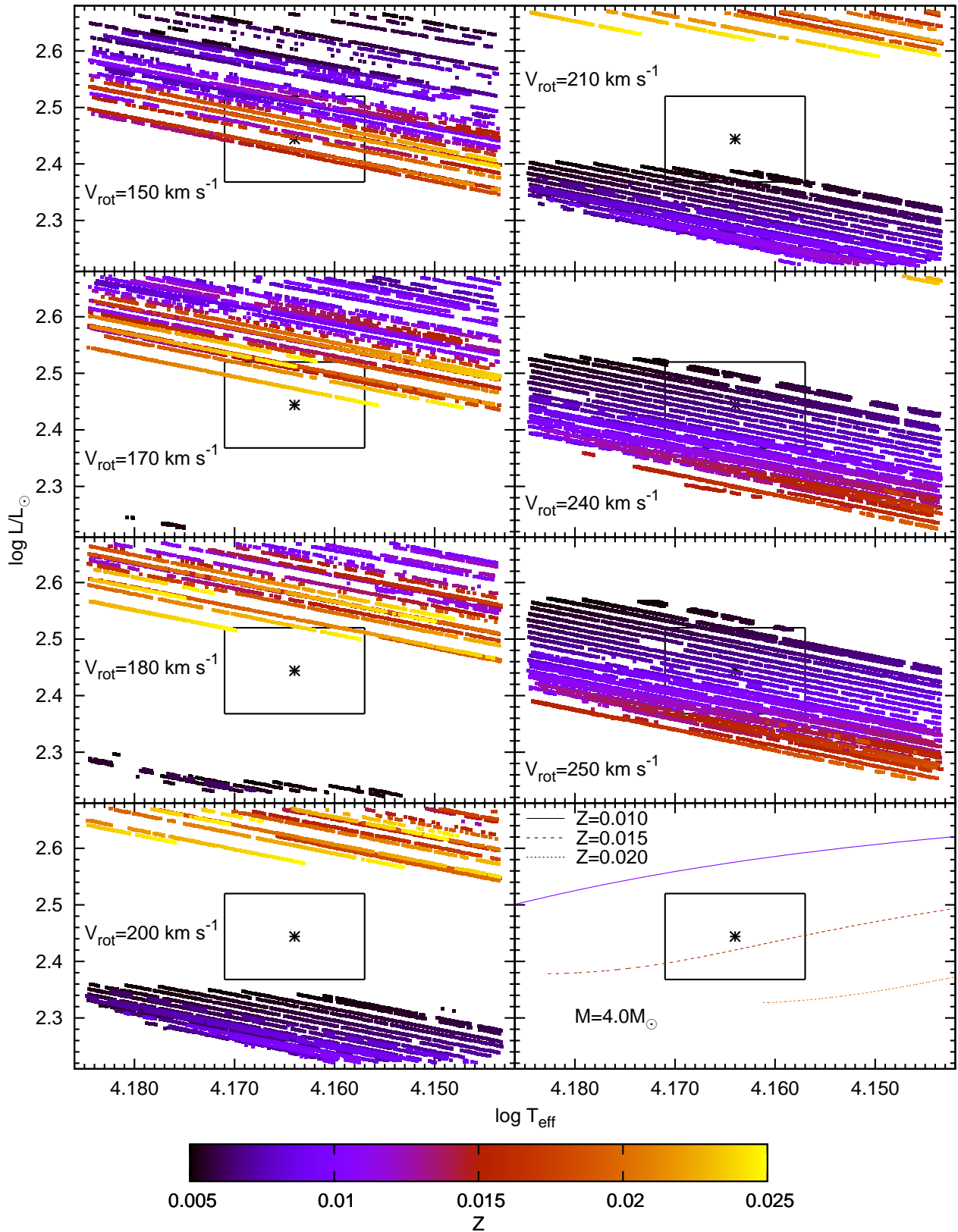}
\caption{The HR diagrams with the models from the G1 grid which fit $\nu_1$ and $\nu_2$, marked, for seven values
         of the rotational velocity, $\mathrm V_{\rm rot}$. Both, stable and unstable modes were plotted.
         In each panel, the inner frame corresponds the $1\sigma$ error box and the entire frame to the $3\sigma$ error box.
         In the bottom right panel there are shown the evolutionary tracks for the three values of metallicity.
            \label{v1v2all_fit_X0_7_ov_0_0_OP}}

\end{figure*}

\begin{figure}
\centering
\includegraphics[angle=-90, width=1\columnwidth]{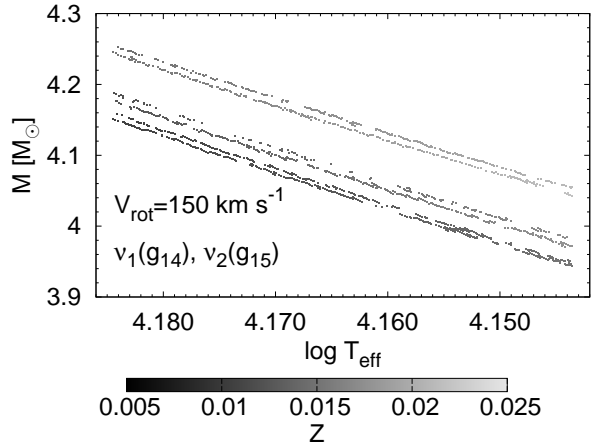}
\caption{Models fitting $\nu_1$ and $\nu_2$ on the diagram $\log T_\mathrm{eff}$ $vs.$ $M$.
        There are shown models from the G1 grid with $\mathrm {V_{rot}}=150$ km s$^{-1}$
        and in which the modes $\mathrm g_{14}$ and $\mathrm g_{15}$ fit the two observed frequencies of HD\,21071.
            \label{v1v2_T_v_M}}

\end{figure}

\begin{figure*}
\centering
\includegraphics[angle=-90, width=2\columnwidth]{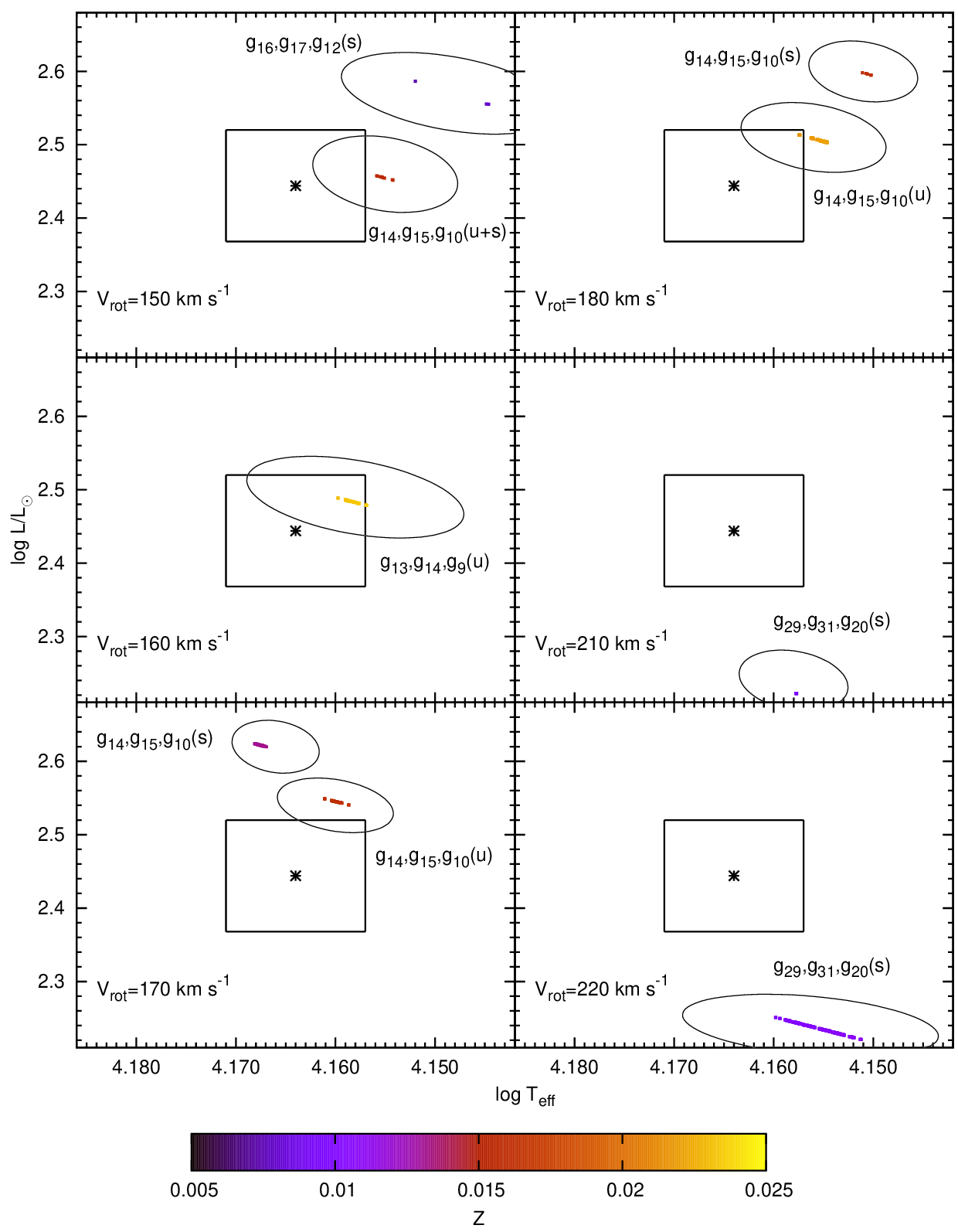}
\caption{The same as in Fig.\,\ref{v1v2all_fit_X0_7_ov_0_0_OP} but models
        fitting the three frequencies $\nu_1$, $\nu_2$ and $\nu_3$ of HD\,21071 are shown.
            \label{v1v2v3_fit_part1}}

\end{figure*}

In Fig.\,\ref{v1v2all_fit_X0_7_ov_0_0_OP}, we put seismic models fitting
$\nu_1$ and $\nu_2$ in the HR diagram for the seven values of $\mathrm V_{\rm rot}$.
The colours are assigned to the values of metallicity, $Z$.
In this figure all modes, stable and unstable, were plotted.
The use of the instability condition significantly reduces the number of seismic models.
{Our seismic modelling showed that for $\mathrm{V_{rot}} \gtrsim 220$ km s$^{-1}$ there are very
few models with unstable modes which reproduce $\nu_1$ and $\nu_2$.
Moreover, these models have rather high metallicity. For example, in the case
of $\mathrm{V_{rot}}=250$ km s$^{-1}$ unstable modes reproducing $\nu_1$ and $\nu_2$
were found in models with $Z> 0.02$.
On the other hand, in the case of the lowest considered rotational velocity, ie.,
$\mathrm{V_{rot}}=150$ km s$^{-1}$ unstable modes reproducing $\nu_1$ and $\nu_2$
were found in models with metallicity as low as $Z \approx 0.0054$.
}

{All our seismic models of HD21071 can be divided into two families.}
%As one can see, there are two families of solutions.
The first one (hereafter F1), associated with the radial orders from $n=13$ to $n=19$,
has on average higher metallicity and is more luminous.
The second one (hereafter F2), associated with the radial orders from $n=26$ to $n=31$,
has on average lower metallicity and is less luminous.
In both cases the models which fit $\nu_1$ and $\nu_2$ become more luminous with increasing rotation velocity.
For the lowest rates of rotation, i.\,e., $\mathrm{V_{rot}}=150$
and 160 km s$^{-1}$, we found only models with the F1 solution within 3$\sigma$ error box.
The seismic models from the F2 solution appear close to the lowest
edge of the 3$\sigma$ error box ($\log L/L_{\sun}\approx 2.22$) from $\mathrm{V_{\rm rot}}=170$ km s$^{-1}$.
With increasing $\mathrm{V_{rot}}$,
we obtained more and more models from F2 and less from F1 in the
space of parameters we considered. For the highest
rotational velocity, $\mathrm{V_{\rm rot}}=250$ km s$^{-1}$, we found only seismic models from F2.
In both cases, there is a clear trend: the lower metallicity  the higher luminosity.

The seismic models of HD\,21071 lie along sloped lines in the HR diagram.
For one pair of the radial orders we usually have a few
nearly parallel sequences of such models differing in metallicity.
A similar structure occurs in the diagram $M~vs.~\log T_{\rm eff}$.
This is because the frequencies of the g modes in the
range of parameters and frequency we consider, in general, decrease with decreasing effective temperature and increasing mass.
The dependence on $Z$ is varied, for the higher radial orders, $n>10$,
the pulsational frequency is an increasing function of $Z$, whereas for $n<10$
there is an oscillatory character of $\nu(Z)$.
In Fig.\,\ref{v1v2_T_v_M}, for clarity, we show the diagram $M~vs.~\log T_{\rm eff}$ for only one pair
of the radial orders, g$_{14}$ and g$_{15}$, and one value of the rotational velocity.

Gaps which sometimes appear along the lines of seismic models
result from the adopted steps in $M,~Z,~T_{\rm eff}$ and $\mathrm V_{\rm rot}$,
which, although small, can cause the omission of some models.

It is worth to notice that in the case of F1  we have always consecutive
radial orders whereas in the F2 solution we have every second radial order.
This is because with increasing $\mathrm{V_{rot}}$ the eigenvalue $\lambda$ of the $\ell=1$, $m=0$ mode
increases and as a result of denser oscillation spectrum of the high radial order modes
is shifted towards higher frequencies. This explains why we have more models from F1 than F2
for lower rotation rates and vice versa.
On the other hand coexistence solutions from F1 and F2 for fixed
rotational velocity (e.\,g., $\mathrm{V_{rot}}=200$ km s$^{-1}$) is associated
with considerably different parameters of the models from both families
of solutions (mainly metallicity and mass) as described in previous paragraph.

{An interesting fact is that most unstable modes comes from
the F1 seismic models. The only exception is for $\mathrm{V_{rot}}=250$ km s$^{-1}$
where some modes from the F2 solution are unstable. With} increasing $\mathrm{V_{rot}}$
there are less and less seismic models with unstable modes fitting the two frequencies $\nu_1$ and $\nu_2$.
On the one hand this is due to lower metallicity of the F2 models which dominate for the higher rotational velocities,
on the other, for higher $\mathrm{V_{rot}}$ the instability domain is shifted towards higher frequencies
than the observed ones.

In the next step we selected models which fit the third observational frequency, $\nu_3$,
which also corresponds to a dipole axisymmetric mode.
Models fitting $\nu_1$, $\nu_2$ and $\nu_3$ are shown in Fig.\,\ref{v1v2v3_fit_part1}.
We can see a further reduction in the number of models.
Moreover, we did not find a solution for each value of $\mathrm{V_{rot}}$.
The radial orders of modes which fit $\nu_1$, $\nu_2$ and $\nu_3$ from the G1 grid
are given in the second column of Table \ref{radial_orders_v1v2v3}.
Unstable solutions exist only for $V_{\rm rot}\le 180$ km s$^{-1}$.

\subsection{Effects of hydrogen abundance, core overshooting and opacities}

There are many input parameters, both, from a model and microphysics, that may
affect pulsational frequencies. Here, we examine the influence of the initial hydrogen abundance, $X_0$,
overshooting from the convective core (in terms of the parameter $\alpha_\mathrm{ov}$) and the opacities.
The effect of the core overshooting was included according to the formulation of \citet{Dziembowski_Pamyatnykh2008}
which takes into account, both, the distance of the overshooting and partial mixing in the overshoot layer.

To this end we constructed three additional grids of models in the same way as explained in Section\,4.2.
In the second grid, G2, we used $X_0=0.75$ (comparing to $X_0=0.70$ in G1),
in the third grid, G3, we added the core overshooting,  $\alpha_\mathrm{ov}=0.2$,
and in the fourth one, G4, we used OPAL \citep{Iglesias1996} opacity tables (instead of OP used in G1).

We found that independently of the adopted grid, all trends noted in G1 are also
present in G2, G3 and G4. Firstly, there are always the two families of solutions.
The first one, F1, with the radial orders of about $n=15$ and the second one, F2,
with the radial orders of about $n=30$. %The radial orders of modes which
%fit $\nu_1$ and $\nu_2$ for all grids are listed in Table\,\ref{radial_orders_v1v2}.

\begin{figure*}
\centering
\includegraphics[angle=-90, width=2\columnwidth]{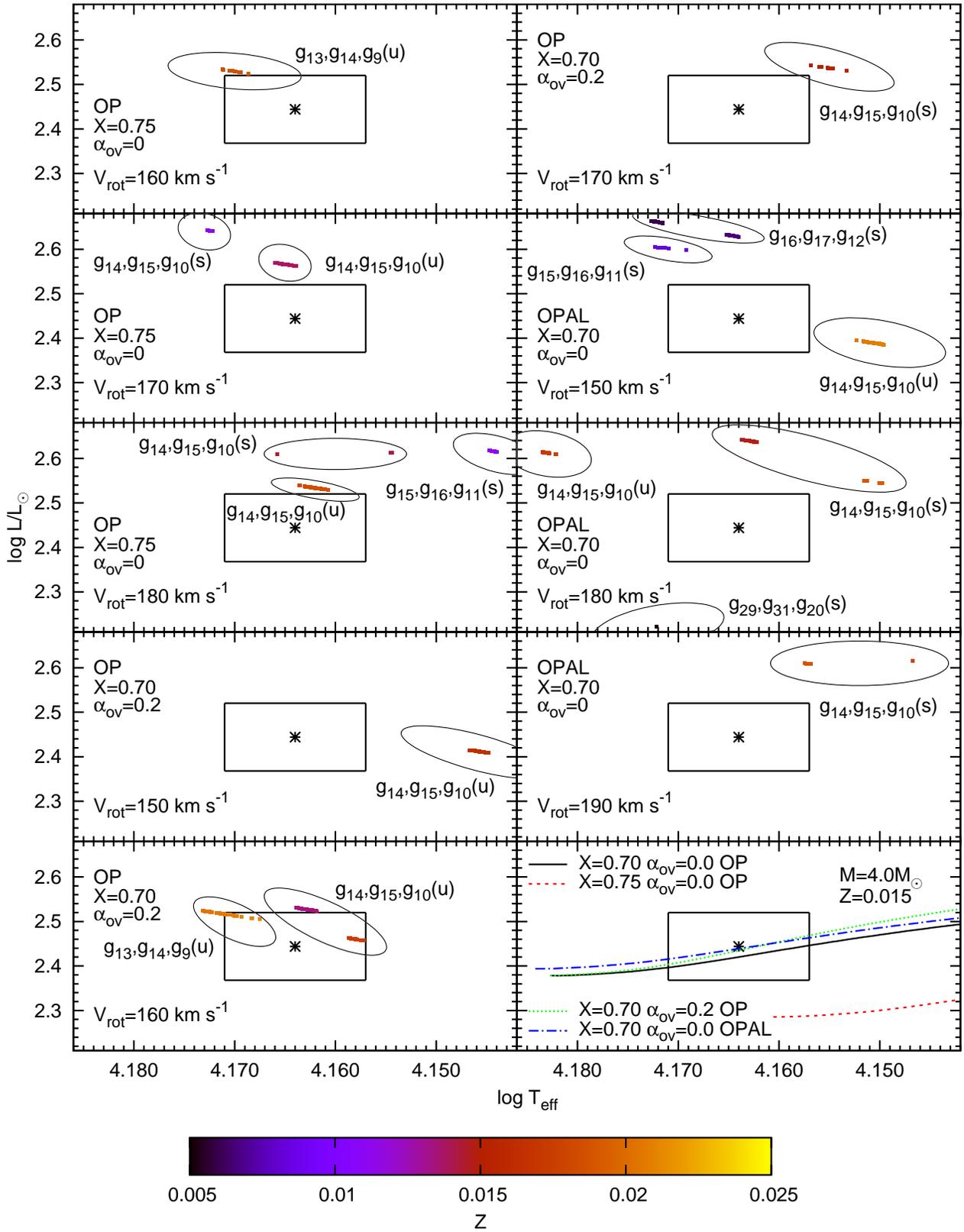}
\caption{The same as in Fig.\,\ref{v1v2v3_fit_part1} but selected models from the grids G2, G3 and G4, are shown.
            \label{v1v2v3_fit_part2}}

\end{figure*}

With an increased abundance of hydrogen (the G2 grid),
we found less seismic models compared to G1, in particular for the F2 solution.

Seismic models with and without overshooting
from the convective core (G3 $vs.$ G1) lie approximately in the same places
of the HR diagram but the G3 models are confined to slightly narrower bands.
A similar picture emerges when we considered seismic models only with unstable modes.

Changing the opacity tables (G4 $vs$. G1) has very little effect on the position
of seismic models in the HR diagram. The G4 models are only slightly more luminous.
The same is true if only unstable modes are considered.
The exception is for seismic models with $\mathrm{V_{rot}}=240$ km s$^{-1}$.
In this case, unstable modes occurs only in the F2 solution.
This is opposite to seismic models calculated with the OP tables.
There are fewer seismic models with unstable modes obtained with the use of
the OPAL opacities. This result is not surprising as it is well known that
for SPB models the computations with the OP tables give more unstable modes
than those with the OPAL ones \citep[e.g.][]{Pamyatnykh1999}. %(e.g., Pamyatnykh 1999).

Models which fit the three frequencies $\nu_1$, $\nu_2$ and $\nu_3$
in the grids G2, G3 and G4 are presented in Fig.\,\ref{v1v2v3_fit_part2}.
In Table\,\ref{radial_orders_v1v2v3}, we list the allowed combinations of the radial orders.

As in the case of G1, only for the lower rotation velocity, $\mathrm{V_{rot}}\le 180$ km s$^{-1}$,
we were able to find models which fit the three frequencies with the instability condition fulfilled.
For $\mathrm{V_{rot}}\ge 190$ km s$^{-1}$, the number of these seismic models is significantly lower
and the modes are stable (cf. Table\,\ref{radial_orders_v1v2v3}).

\begin{table*}
\centering
\begin{minipage}{\textwidth}
  \caption{Ranges of the astrophysical parameters  for the groups of seismic models
           (rows indicated by `grp' in the sixth column) and the parameters of the selected models
           for each group (rows indicated by `rep' in the sixth column).
           All models reproduce the three observed frequencies with the unstable dipole axisymmetric modes.
           Groups of models are defined as models with similar parameters
           forming distinct groups in Figs.\,\ref{v1v2v3_fit_part1} and \ref{v1v2v3_fit_part2}.
           In the last column the goodness of fit is given for representative models (cf. Eq.\,\ref{good_fit}).}
\label{models}
\centering
  \begin{tabular}{ccccccccccc}
  \hline
Grid & opacity  & $X_0$ & $\alpha_\mathrm{ov}$ & $V_\mathrm{rot}$ & ID & $Z$      & $M$ ($\mathrm M_{\sun}$)  & $\log T_\mathrm{eff}$ &  $\log L/\mathrm L_{\sun}$  &      $d$ \\
     &          &       &                      &  (km s$^{-1}$)   &    &  &     &        &    &       \\

\hline\hline
G1  & OP   & 0.70  &   0.0    &  150   & grp a  & $0.0157 - 0.0160$ & $4.031 - 4.039$ & $4.1542 - 4.1559$  & $2.452 - 2.458$ & -- \\
G1  & OP   & 0.70  &   0.0    &  150   & rep a  &  0.01575          &  4.0357         &  4.15528           & 2.4554          &  0.139 \\
G1  & OP   & 0.70  &   0.0    &  160   & grp b  & $0.0231 - 0.0236$ & $4.372 - 4.388$ & $4.1568 - 4.1597$  & $2.479 - 2.489$ & -- \\
G1  & OP   & 0.70  &   0.0    &  160   & rep b  & 0.02340           & 4.3793          & 4.15814            & 2.4831          &  0.331  \\
G1  & OP   & 0.70  &   0.0    &  170   & grp c  & $0.0161 - 0.0164$ & $4.218 - 4.232$ & $4.1587 - 4.1611$  & $2.541 - 2.549$ & -- \\
G1  & OP   & 0.70  &   0.0    &  170   & rep c  & 0.01630           & 4.2221          & 4.15948            & 2.5435          &   0.312 \\
G1  & OP   & 0.70  &   0.0    &  180   & grp d  & $0.0216 - 0.0222$ & $4.351 - 4.364$ & $4.1547 - 4.1574$  & $2.503 - 2.513$ & -- \\
%G1  & OP   & 0.70  &   0.0    &  180   & rep d  & 0.02205           & 4.3543          & 4.15527            & 2.5054          & 0.247 \\
G1  & OP   & 0.70  &   0.0    &  180   & rep d  & 0.02210           & 4.3521          & 4.15492            & 2.5042          & 0.624 \\
\hline
G2  & OP   & 0.75  &   0.0    &  160   & grp e  & $0.0184 - 0.0188$ & $4.647 - 4.661$ & $4.1686 - 4.1712$  & $2.524 - 2.534$ & -- \\
G2  & OP   & 0.75  &   0.0    &  160   & rep e  & 0.01855           & 4.6566          & 4.17028            & 2.5304          & 0.230 \\
G2  & OP   & 0.75  &   0.0    &  170   & grp f  & $0.0136 - 0.0139$ & $4.455 - 4.469$ & $4.1639 - 4.1660$  & $2.562 - 2.569$ & -- \\
G2  & OP   & 0.75  &   0.0    &  170   & rep f  & 0.01375           & 4.4643          & 4.16493            & 2.5657          & 0.720 \\
G2  & OP   & 0.75  &   0.0    &  180   & grp g  & $0.0181 - 0.0186$ & $4.595 - 4.612$ & $4.1607 - 4.1636$  & $2.529 - 2.539$ & -- \\
G2  & OP   & 0.75  &   0.0    &  180   & rep g  & 0.01835           & 4.6040          & 4.16237            & 2.5347          & 0.003  \\
\hline
G3  & OP   & 0.70  &   0.2    &  150   & grp h  & $0.0163 - 0.0165$ & $3.909 - 3.919$ & $4.1448 - 4.1467$  & $2.409 - 2.415$ & -- \\
G3  & OP   & 0.70  &   0.2    &  150   & rep h  & 0.01640           & 3.9139          & 4.14568            & 2.4114          & 0.740 \\
G3  & OP   & 0.70  &   0.2    &  160   & grp i  & $0.0132 - 0.0135$ & $4.035 - 4.045$ & $4.1619 - 4.1639$  & $2.523 - 2.531$ & -- \\
%G3  & OP   & 0.70  &   0.2    &  160   & rep i  & 0.01340           & 4.0360          & 4.16242            & 2.5254          & 0.288 \\
G3  & OP   & 0.70  &   0.2    &  160   & rep i  & 0.01340           & 4.0365          & 4.16247            & 2.5255          & 0.497 \\
G3  & OP   & 0.70  &   0.2    &  160   & grp j  & $0.0173 - 0.0176$ & $4.097 - 4.106$ & $4.1571 - 4.1587$  & $2.457 - 2.463$ & -- \\
G3  & OP   & 0.70  &   0.2    &  160   & rep j  & 0.01740           & 4.0997          & 4.15804            & 2.4604          & 0.214 \\
G3  & OP   & 0.70  &   0.2    &  160   & grp k  & $0.0203 - 0.0213$ & $4.382 - 4.412$ & $4.1675 - 4.1732$  & $2.504 - 2.524$ & -- \\
%G3  & OP   & 0.70  &   0.2    &  160   & rep k  & 0.02090           & 4.3937          & 4.16975            & 2.5122          & 0.052 \\
G3  & OP   & 0.70  &   0.2    &  160   & rep k  & 0.02115           & 4.3860          & 4.16833            & 2.5072          & 0.326 \\
\hline
G4  & OPAL & 0.70  &   0.2    &  150   & grp l  & $0.0204 - 0.0209$ & $4.058 - 4.071$ & $4.1496 - 4.1523$  & $2.386 - 2.395$ & -- \\
G4  & OPAL & 0.70  &   0.2    &  150   & rep l  & 0.02075           & 4.0651          & 4.15081            & 2.3900          & 0.413 \\
G4  & OPAL & 0.70  &   0.2    &  180   & grp m  & $0.0171 - 0.0175$ & $4.484 - 4.489$ & $4.1821 - 4.1835$  & $2.609 - 2.614$ & -- \\
%G4  & OPAL & 0.70  &   0.2    &  180   & rep m  & 0.01745           & 4.4845          & 4.18216            & 2.6091          & 0.851 \\
G4  & OPAL & 0.70  &   0.2    &  180   & rep m  & 0.01745           & 4.4846          & 4.18217            & 2.6091          & 0.433 \\
\hline
\end{tabular}
\end{minipage}
\end{table*}

\begin{figure*}
\centering
\includegraphics[angle=-90, width=2\columnwidth]{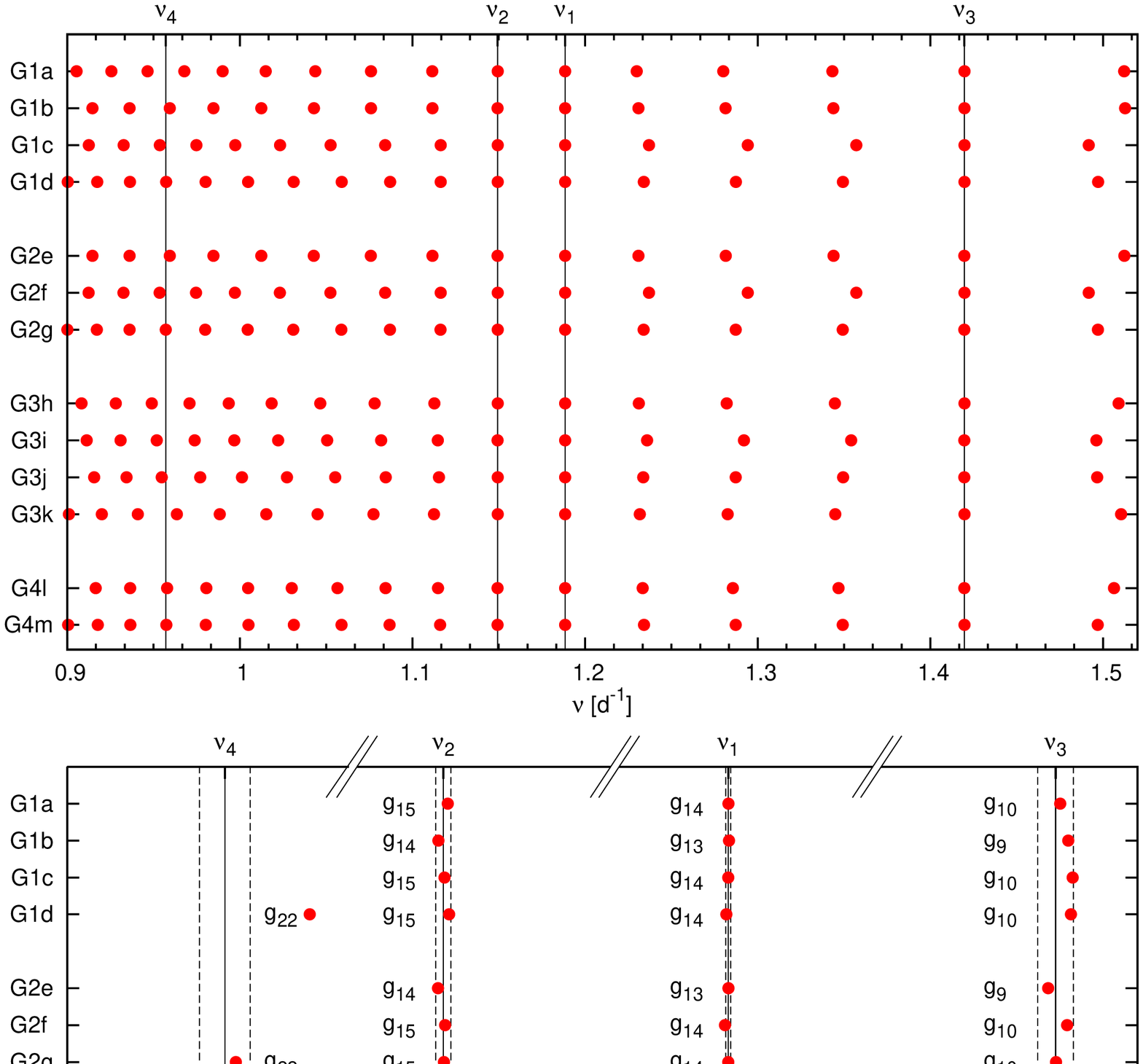}
\caption{The frequencies of dipole axisymmetric modes in the representative models (see Table\,\ref{models})
         which reproduce three well identified frequencies, $\nu_1$, $\nu_2$ and $\nu_3$, of HD\,21071.
         A zoom-in on the frequencies around the observed values is shown in the bottom panel for clarity.
         In this panel the frequency of the quadrupole retrograde mode ($\ell=2,~m=-2$) is added as an inverted triangle.
         The vertical solid lines indicate the observed frequencies and the vertical dashed lines, the observational errors of the frequencies.
        \label{HD21071_modelsv3}}

\end{figure*}

To summarize our results of seismic modelling of the star HD\,21071,
in Table\,\ref{models} we give the ranges of the stellar parameters of models with unstable modes which fit
the three frequencies, $\nu_1$, $\nu_2$ and $\nu_3$, for the four grids.
The parameters are provided for each separate group of models with similar properties,
shown in Figs.\,\ref{v1v2v3_fit_part1} and \ref{v1v2v3_fit_part2} by ovals.
The groups are marked as 'grp' in Table\,\ref{models}. In addition, we give also the  parameters
of representative models for each group (marked as 'rep' in Table\,\ref{models}).
In the last column we give, for the representative seismic models, the goodness of the fit defined as
\begin{equation}
\label{good_fit}
d= \frac{1}{3} \sum_{n=1}^3  \frac{\left( \nu_n^\mathrm o - \nu_n^\mathrm t\right)^2}{\sigma^2_n} ,
\end{equation}
where $\nu_n^\mathrm o$ and $\nu_n^\mathrm t$ are observed and theoretical frequency, respectively,
and $\sigma_n$ are the observational errors of the frequencies.

{It should be mentioned that masses of seismic models from Table\,\ref{models}
are in the range $M\in\left<3.909,4.661\right> \mathrm M_{\sun}$ whereas masses of evolutionary models
which fall into the 3$\sigma$ error box are in a slightly wider range $ M\in\left<3.04,4.99\right>\mathrm M_{\sun}$.
The range of masses of evolutionary models given above is wider than the one given in Section 2
because here we take into account models with all considered metallicities ($Z\in\left<0.005,0.025\right>$) and from
all grids.}

In Fig.\,\ref{HD21071_modelsv3}, we compare the oscillation spectrum of HD\,21071
with the theoretical spectra corresponding to the representative seismic models from Table\,\ref{models}.
In the bottom panel of Fig.\,\ref{HD21071_modelsv3}, we show a zoom-in on the frequencies around the observed values.
As one can see, the model G2g reproduces also the frequency $\nu_4$ within the observational errors
if it is the mode $\ell=1~,m=0$, g$_{22}$, a possibility consistent with our mode identifications.
Similarly, the quadrupole retrograde mode $\ell=2,~m=-2$, g$_{38}$ in the model G3i fits the frequency $\nu_4$.
This mode is also allowed by our mode identifications.

We can also see that the frequency of the mode $\ell=1~,m=0$, g$_{22}$ is close to $\nu_4$ in the models G1d and G4m.
In a similar distance is also the mode, $\ell=2,~m=-2$, g$_{31}$, in the model G3k.
Given the numerical accuracy of theoretical frequencies, which is about $\Delta\nu=0.0001$ d$^{-1}$,
we have to accept also these solutions.
The important result is that these identifications are valid for the all seismic models constructed in this paper.

\section{Conclusions}

This paper focused on the challenges and prospects for seismic modelling of slowly pulsating B-type stars
with the effects of rotation taken into account.
We used as an example the star HD\,21071 which pulsates in four frequencies of which the three have
a unique mode identification.
Firstly, we showed that non-rotating models cannot account for the two highest amplitude frequencies, $\nu_1=1.18843$
and $\nu_2=1.14934$ d$^{-1}$, and including the effects of rotation on high-order g modes is indispensable.
Then, having unambiguous determination of the two angular numbers $(\ell,~m)$ for the three observed frequencies
of HD\,21071 ($\nu_1$, $\nu_2$ and $\nu_3=1.41968$ d$^{-1}$), and constraints on the range of the rotational velocity, $\mathrm{V_{rot}}\in\left<150,\,250\right>$ km s$^{-1}$, we constructed rotating seismic models which reproduce these three frequencies. We examined the effects of the initial abundance of
hydrogen, an amount of the core overshooting and the opacity data, considering four grids of parameters.
Due to the high density of the theoretical oscillation spectra, a large number of solutions have been obtained.
Despite of that, only two combinations of the radial orders, $n$, were allowed:
one set is around $n=15$ and the second one around $n=30$.
Moreover, the instability condition reduced the number of seismic models significantly.
Therefore, accurate calculations of the opacity data are of the utmost importance.

Among seismic models fitting the three frequencies, we found some that reproduce also the forth frequency, $\nu_4=0.95706$ d$^{-1}$.
In all grids of models only the dipole axisymmetric modes g$_{22}$ or quadrupole modes $\ell=2,~m=-2$, g$_{38}$, g$_{31}$ have frequencies
close to $\nu_4$ (given the observational errors and numerical accuracy). Both modes are in agreement with the mode identification.
Fitting the forth frequency further reduced the number of seismic models. However, to derive more stringent constraints on
opacities, core overshooting etc. more well identified frequencies are necessary.
Because we tried to constrain seven parameters ($M,~T_{\rm eff},~ X_0,~Z,~\kappa,~\alpha_{\rm ov},~\mathrm V_{\rm rot}$),
we need at least seven well identified pulsational frequencies.

There are many SPB stars with far more rich oscillation spectra obtained from the space missions,
like MOST, CoRoT and Kepler, but the problem is their mode identification.
Even if there are some patterns in the oscillation spectra, they can be accidental, i.e., composed of modes with different pairs ($\ell,~m$),
and cannot be interpreted in term of the asymptotic theory \citep{WSJDDWD2014}.
Therefore, except for the excellent example of KIC\,10526294 with 19 consecutive rotationally split dipole modes \citep{Papics2014},
there is no other reliable seismic modelling of an SPB star.

The aim of this paper was to show that seismic modelling of SPB stars is feasible
even if no asymptotic pattern is observed in the oscillation spectra and a limited number of modes is detected,
provided the angular numbers $(\ell,~m)$ are determined unambiguously
and the effects of rotation are properly included.
{Therefore, a  hope is pinned on the two colour data from the BRITE mission
which, besides more frequencies and better accuracy, can be used to identify the pulsational modes.}

\section*{Acknowledgments}
We gratefully thank Mike Jerzykiewicz for his carefully reading the manuscript.
WS was financially supported by the Polish NCN grant DEC-2012/05/N/ST9/03905 and JDD
by the Polish NCN grants 2011/01/M/ST9/05914,  2011/01/B/ST9/05448.
Calculations have been carried out using resources provided by Wroc{\l}aw Centre
for Networking and Supercomputing (http://wcss.pl), grant No. 265

\appendix
\section{Asymmetry of the rotationally split modes in the framework of the traditional approximation}
\label{app1}

{

Linear theory of pulsation predicts asymmetry between rotationally
split components of the given mode, ie., retrograde modes should be closer
to the axisymmetric ones than prograde modes.
In Table\,\ref{rot_split_model} we list frequencies of dipole modes and the
differences between their components for the model at the centre of the error box
($M=3.69~\mathrm M_{\sun}$, $\log T_\mathrm{eff}=4.164$, $\log L/\mathrm L_{\sun}=2.444$, $X_0 =0.7$, $Z =0.0082$,
$R=2.62~\mathrm R_{\sun}$). The rotation splitting was computed with  the traditional approximation
and the rotational velocities were chosen to approximately reproduce the value of the observed splitting
in HD\,21071.
As one can see, in all cases retrograde modes are closer to axisymmetric modes than prograde ones
and asymmetry is of the order of 0.01 d$^{-1}$.
This property is also clearly seen in Fig.\,\ref{osc_split},
where the observed triplet is compared with the theoretical spectrum
calculated for the model at the centre of error box of HD\,21071.
The observed frequencies were shifted in order to align the central peak, $\nu_1$, with
the $\mathrm g_{12}\left(m=0\right)$ mode.

\begin{table*}
 \centering
 \begin{minipage}{\textwidth}
  \caption{Theoretical frequencies of dipole modes for several radial orders, $n$,
  and the differences between their rotationally split components,
  for the central model of HD\,21071 for the three values of the rotational velocity.}
\label{rot_split_model}
\centering
  \begin{tabular}{ccccccc}
  \hline
$\mathrm{V_{rot}}$  & $n$  & $\nu \left(m=-1\right)$ & $\nu \left(m=0\right)$ & $\nu \left(m=1\right)$ &   $\nu \left(m=0\right) - \nu \left(m=-1\right)$ &   $\nu \left(m=1\right) - \nu \left(m=0\right)$\\
$\left(\mathrm{km\,s^{-1}} \right)$     &      & $\left(\mathrm{d^{-1}} \right)$ & $\left(\mathrm{d^{-1}} \right)$ & $\left(\mathrm{d^{-1}} \right)$ & $\left(\mathrm{d^{-1}} \right)$ & $\left(\mathrm{d^{-1}} \right)$     \\
\hline
\hline
60 & 13 &   0.86855 &   1.08271 &  1.30635 & 0.21416 & 0.22364 \\
   & 12 &   0.93075 &   1.14612 &  1.37041 & 0.21537 & 0.22429 \\
   & 11 &   1.01204 &   1.22871 &  1.45373 & 0.21667 & 0.22502 \\
   & 10 &   1.11148 &   1.32946 &  1.55497 & 0.21798 & 0.22551 \\
\hline
%61 & 13 &   0.86758 &   1.08507 &  1.31226 & 0.21749 & 0.22719 \\
%   & 12 &   0.92967 &   1.14843 &  1.37631 & 0.21876 & 0.22788 \\
%   & 11 &   1.01072 &   1.23080 &  1.45939 & 0.22008 & 0.22859 \\
%   & 10 &   1.10999 &   1.33134 &  1.56058 & 0.22135 & 0.22924 \\
%\hline
62 & 13 &   0.86667 &   1.08742 &  1.31828 & 0.22075 & 0.23086 \\
   & 12 &   0.92855 &   1.15066 &  1.38221 & 0.22211 & 0.23155 \\
   & 11 &   1.00942 &   1.23300 &  1.46530 & 0.22358 & 0.23230 \\
   & 10 &   1.10854 &   1.33333 &  1.56619 & 0.22479 & 0.23286 \\
\hline
%63 & 13 &   0.86582 & 1.08976   &  1.32433 & 0.22394 & 0.23457 \\
%   & 12 &   0.92748 & 1.15284   &  1.38810 & 0.22536 & 0.23526 \\
%   & 11 &   1.00829 & 1.23510   &  1.47102 & 0.22681 & 0.23592 \\
%   & 10 &   1.10706 & 1.33536   &  1.57202 & 0.22830 & 0.23666 \\
%\hline
%64 & 13 &   0.86496 & 1.09224   &  1.33019 & 0.22728 & 0.23795 \\
%   & 12 &   0.92646 & 1.15520   &  1.39398 & 0.22874 & 0.23878 \\
%   & 11 &   1.00704 & 1.23722   &  1.47685 & 0.23018 & 0.23963 \\
%   & 10 &   1.10566 & 1.33738   &  1.57761 & 0.23172 & 0.24023 \\
%\hline
65 & 13 &   0.86408 & 1.09459   &  1.33623 & 0.23051 & 0.24164 \\
   & 12 &   0.92547 & 1.15755   &  1.39986 & 0.23208 & 0.24231 \\
   & 11 &   1.00584 & 1.23941   &  1.48263 & 0.23357 & 0.24322 \\
   & 10 &   1.10431 & 1.33940   &  1.58344 & 0.23509 & 0.24404 \\
\hline
\end{tabular}
\end{minipage}
\end{table*}

\begin{figure}
\centering
\includegraphics[angle=-90, width=\columnwidth]{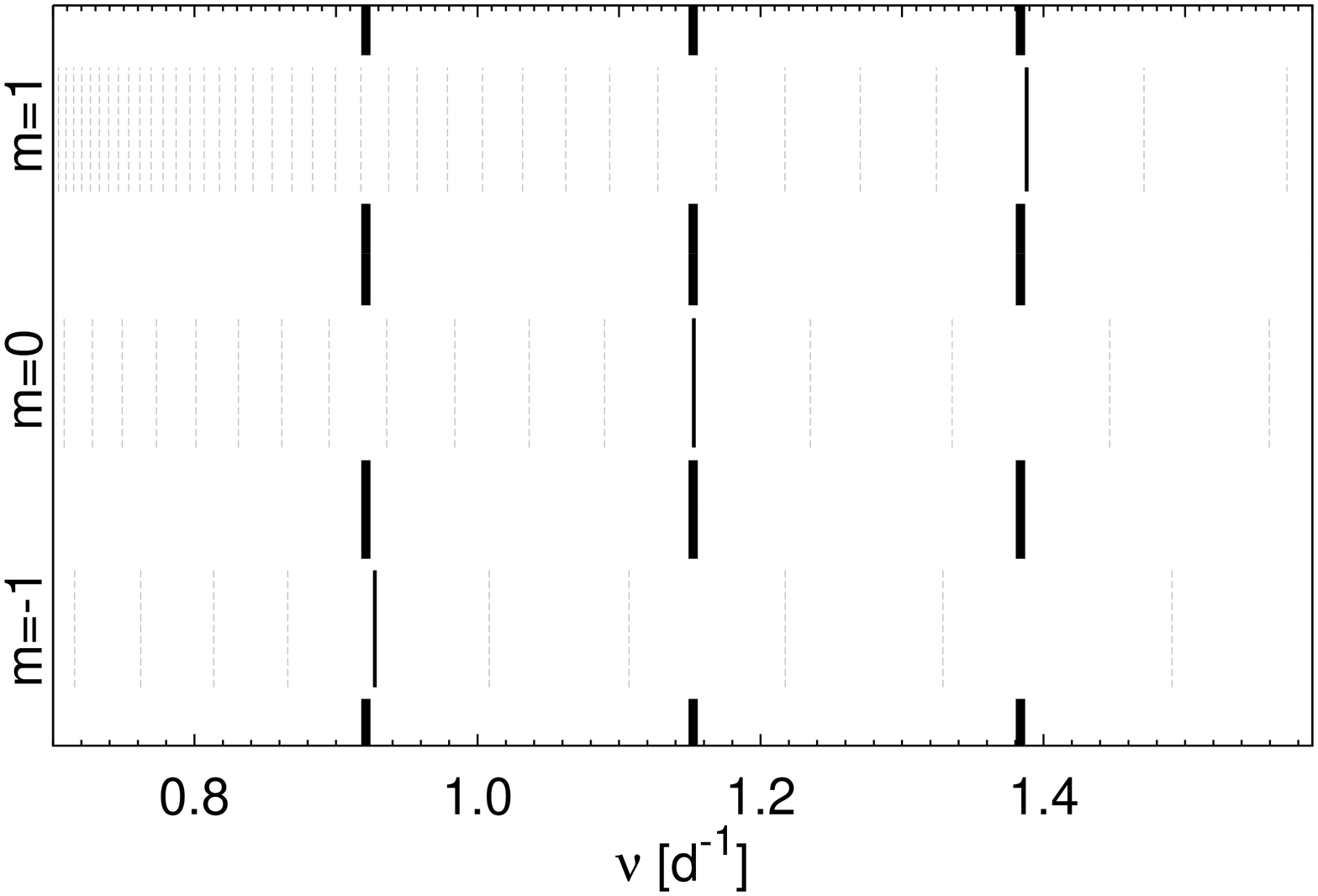}
\caption{The theoretical frequencies (dashed lines) of dipole modes calculated with the traditional approximation for the
         model at the centre of the error box of HD\,21071 with $\mathrm{V_{rot}}=63$ km s$^{-1}$.
         The mode $g_{12}$ is highlighted with a thin solid line. The observed triplet (thick solid lines)
         is shifted by -- 0.036 d$^{-1}$ to align its central frequency with the $g_{12}\left(m=0\right)$ mode.
            \label{osc_split}}
\end{figure}

}
\label{lastpage}

\end{document}